\documentclass[journal=nalefd,manuscript=article]{achemso}
\setkeys{acs}{articletitle = true}

\usepackage[normalem]{ulem}
\usepackage{graphicx}
\usepackage{braket}
\usepackage{amsmath}
\usepackage{bm}
\usepackage{tabu}
\usepackage{subcaption}
\usepackage{caption}
\usepackage{xspace}
\usepackage[section]{placeins}
\usepackage{color}
\usepackage[english]{babel}
\usepackage{natbib}
\usepackage{multirow}

\newcommand{\angstrom}{\mbox{\normalfont\AA}}

\title{Excitonic States in Semiconducting Two-dimensional Perovskites}

\author{Alejandro Molina-S\'anchez}

\affiliation{$^*$Institute of Materials Science (ICMUV), University of Valencia, Catedr\'{a}tico Beltr\'{a}n 2, E-46980, Valencia, Spain}

\email{alejandro.molina@uv.es}

\date{\today}


\begin{document}

\begin{abstract}

Hybrid organic/inorganic perovskites have emerged as efficient semiconductor materials
for applications in photo-voltaic solar cells with conversion efficiency above
20 \%. Recent experiments have synthesized ultra-thin two-dimensional (2D) organic
perovskites with optical properties similar to those of 2D
materials like monolayer MoS$_2$: large exciton binding energy and excitonic
effects at room temperature. In addition, 2D perovskites are synthesized with a simple 
fabrication process with potential low-cost and large-scale manufacture.

Up to now, state-of-the-art simulations of the excitonic states have been limited
to the study of bulk organic perovskites. The large number of atoms in the unit cell and
the complex role of the organic molecules makes inefficient the use
of \textit{ab initio} methods. In this work, we define a simplified crystal
structure to calculate the
optical properties of 2D perovskites, replacing the molecular
cations with inorganic atoms. We can thus apply
state-of-the-art, parameter-free and predictive \textit{ab initio} methods like the GW method and the 
Bethe-Salpeter equation to obtain the excitonic states
of a model 2D perovskite. We find that optical properties of 2D perovskites are strongly influenced by excitonic effects, with
binding energies up to 600 meV. Moreover, the optical absorption is carried out at the bromine and lead atoms and therefore
the results are useful for a qualitatively understanding of the optical
properties of organic 2D perovskites. 

\end{abstract}

\maketitle

\section{Introduction}

Semiconducting hybrid organic-inorganic perovskites (HOPV) and all-inorganic
perovskites (AIP) are key materials for
light harvesting and solar cells
applications,\cite{Graetzel2014,Ahmad2015,Yettapu2016,Frost2016,Kovalenko2017} with conversion 
efficiencies reported above 20\%\cite{Bi2016,Yang2017}. Perovskites can also be
synthetized in layered, two-dimensional (2D) structures,\cite{Dou2015,Quan2016} in analogy with the
well-known 2D materials like graphene or monolayer
MoS$_2$.\cite{Geim2013,Castellanos-Gomez2016} An interesting 2D perovskite
is the Ruddlesden-Popper perovskite, defined by the formula
A$_{n+1}$B$_n$X$_{3n+1}$ where $n$ is the number of layers and A is usually
an organic molecule like CH$_3$NH$_3$PbX$_3$, B stays for lead and X for bromine or
iodine.\cite{Mao2018,Stoumpos2016,Soe2018} They 
exhibit intense room-temperature photoluminescence,
significant flexibility and strong quantum
confinement.\cite{Traore2018,Tsai2018,Blancon2017} 


Moreover, monolayer 2D perovskites ($n=1$) have other interesting optical
properties similar to those of 2D semiconductors
like hexagonal BN or monolayer
MoS$_2$.\cite{Splendiani2010,Hill2015,Molina-Sanchez2015,Cassabois2016,Galvani2016,Torun2018}
We expect a large excitonic binding energy due to the dramatic reduction
of the dielectric screening.\cite{Chernikov2014,Latini2015} For instance,
the reported experimental value of the exciton binding energy is 467 meV
for 2D perovskites (monolayer
(CH$_3$NH$_3$)$_2$PbI$_4$),\cite{Blancon2018a} 
corroborated with semi-empirical
Bethe-Salpeter Equation calculations.\cite{Pedesseau2016} In 2D perovskites 
of (C$_{6}$H$_{13}$NH$_3$)$_2$PbI$_4$ the exciton binding energy was
estimated in 367 meV\cite{Tanaka2005}. Experiments have also reported 
an unusual thickness dependence of the exciton
characteristics.\cite{Blancon2018,Blancon2018a}

In addition to the remarkable optical and excitonic properties, 2D HOPV have
others advantages with respect other 2D materials. For example, 
optical properties like the absorption or emission can be modulated by
the choosing the composition (by exchanging bromine by iodine, or by using tin 
in place of lead).\cite{Zhou2017,Mao2018,Sapori2016} Moreover, using appropriate molecules
as spacers is possible to grow few-layer 2D perovskites with a
chosen number of layers.\cite{Sapori2016} 


In general, accurate theoretical description of perovskites requires
the use of large unit cells to capture the disorder related to organic
cations.\cite{Mosconi2013,Even2013} Most theoretical contributions use
density functional theory (DFT) simulations,\cite{Gebhardt2017} with
focus on the electronic and optical properties of bulk HOPV\cite{Huang2013,Huang2015}
and single-crystal perovskites.\cite{Park2015} The large number of
atoms in the unit cell of organic 2D perovskites
makes prohibitive the simulations including many-body effects such as
the Bethe-Salpeter equation (BSE) to account for the excitonic
effects.\cite{Pedesseau2016,Rodriguez-Romero2017} Only the studies of bulk
perovskites with a few atoms
in the unit cell permit the use of fully \textit{ab initio} approaches such as the GW
method or \textit{ab initio} BSE.\cite{Filip2015} The current alternatives to 
fully \textit{ab initio} simulations are semi-empirical atomistic models 
of the dielectric constant.\cite{Giustino2003} They can investigate
large systems very efficiently and provide an useful insight 
based on solid-state physics concepts, but they lack the accuracy and
predictive power of \textit{ab initio} BSE.\cite{Even2015}


In spite of the versatility of the atomistic models the computation of the
excitonic states within a full \textit{ab initio} framework is desirable to obtain
predictive information. We propose to use a
simplified 2D monolayer perovskite structure, in which the use of
\textit{ab initio} approach becomes efficient and reliable.
We replace the organic molecule in the 2D
perovskite by a cation atom (cesium in our case),
obtaining the layered crystal Cs$_{n+1}$Pb$_n$Br$_{3n+1}$,
where $n$ is the number of layers. The monolayer structure ($n=1$)
Cs$_{2}$PbBr$_{4}$ is a suitable system to apply state-of-the-art ab initio
methods. We can reproduce the conditions of reduced dielectric
screening to obtain the excitonic states and to simulate the optical 
properties.\cite{Latini2015} Moreover, our \textit{ab initio}
calculations include the spin-orbit interaction,\cite{Even2013} necessary to obtain
reliable excitonic states, quantitative exciton binding energies, and to 
describe realistically the optical properties.

We find remarkable excitonic effects on the optical properties of monolayer 2D
perovskites, with exciton binding at the order of 0.5 eV, the existence
of dark excitons below the optical absorption edge, and a strong anisotropy
of the optical response with respect to the light polarization. Moreover, our
simplified 2D crystal can also be used to investigate the role of the number of
layers or the different chemical composition.

\section{Electronic structure}

\begin{figure}
\includegraphics[width=7 cm]{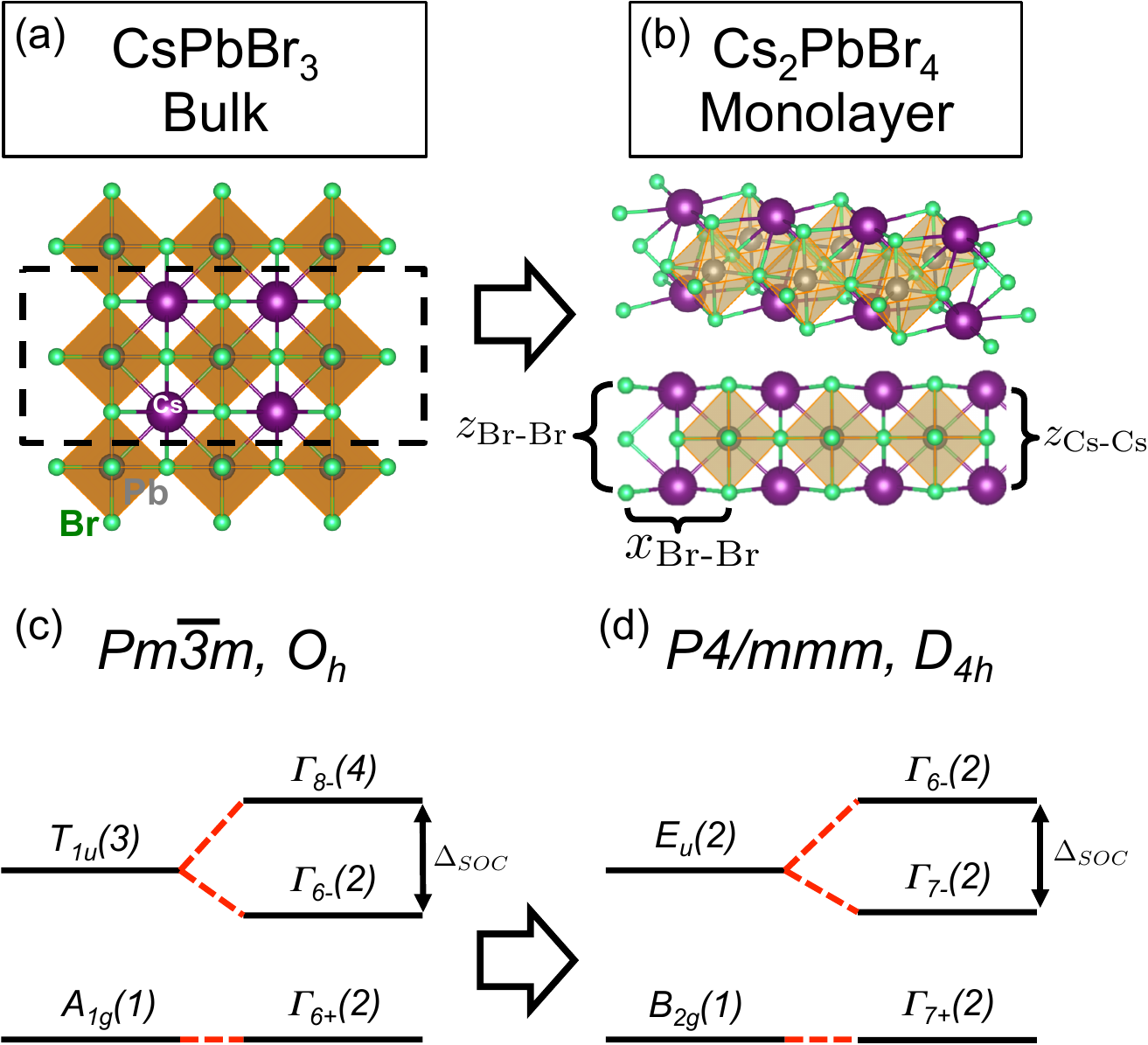}
\caption{Crystal structure of (a) bulk CsPbBr$_{3}$ and (b) monolayer
Cs$_{2}$PbBr$_{4}$. Irreducible representations of the conduction and valence
band states at the bandgap at for (c) bulk and (d) monolayer. The bandgap is
located at the points R and M of the Brillouin zone, respectively.}
\label{crystal}
\end{figure}

The monolayer of inorganic perovskite Cs$_{2}$PbBr$_{4}$ is constructed from the bulk
perovskite CsPbBr$_3$,\cite{Rui2017} as shown in Fig. \ref{crystal} (a). We isolate 
a layer of the octahedron compose by six bromine atoms with one lead atom at the center,
and encapsulated with cesium atoms, as shown in Fig. \ref{crystal} (b),
resulting in a unit cell of 7 atoms. In the
case of bulk, the space and point group symmetry are $Pm\overline{3}m$ and $O_h$,
respectively. The monolayer symmetry has space group 4$/$mmm and point group $D_{4h}$.

The first step of our study is to determine the stability of the monolayer perovskite. We 
have used the DFT implementation of
Quantum Espresso.\cite{Giannozzi2009} We work within the local-density approximation (LDA) with 
norm-conserving fully relativistic pseudopotentials.\cite{Lejaeghere2016}  We
have optimized the crystal structure, by relaxing the in-plane lattice parameter and
the interatomic positions. After relaxation the octahedron formed by bromine
atoms is elongated out-of-plane
(5.88 \angstrom) and compressed in-plane (5.69 \angstrom) and the inter-atomic distances
between cesium atoms are also reduced (from  5.74 to 5.22 \angstrom).
The charge
density is calculated with a $\bf{k}$-sampling of $12 \times 12 \times 1$, energy cutoff of
100 Ry, and a vacuum distance of 18 \AA. Afterwards we have calculated the phonon modes in the
relaxed structure. All phonons have positive frequencies and therefore the structure is
stable (see phonons band structure in Fig. S2 of the supporting information).

\begin{figure*}
\includegraphics[width=14 cm]{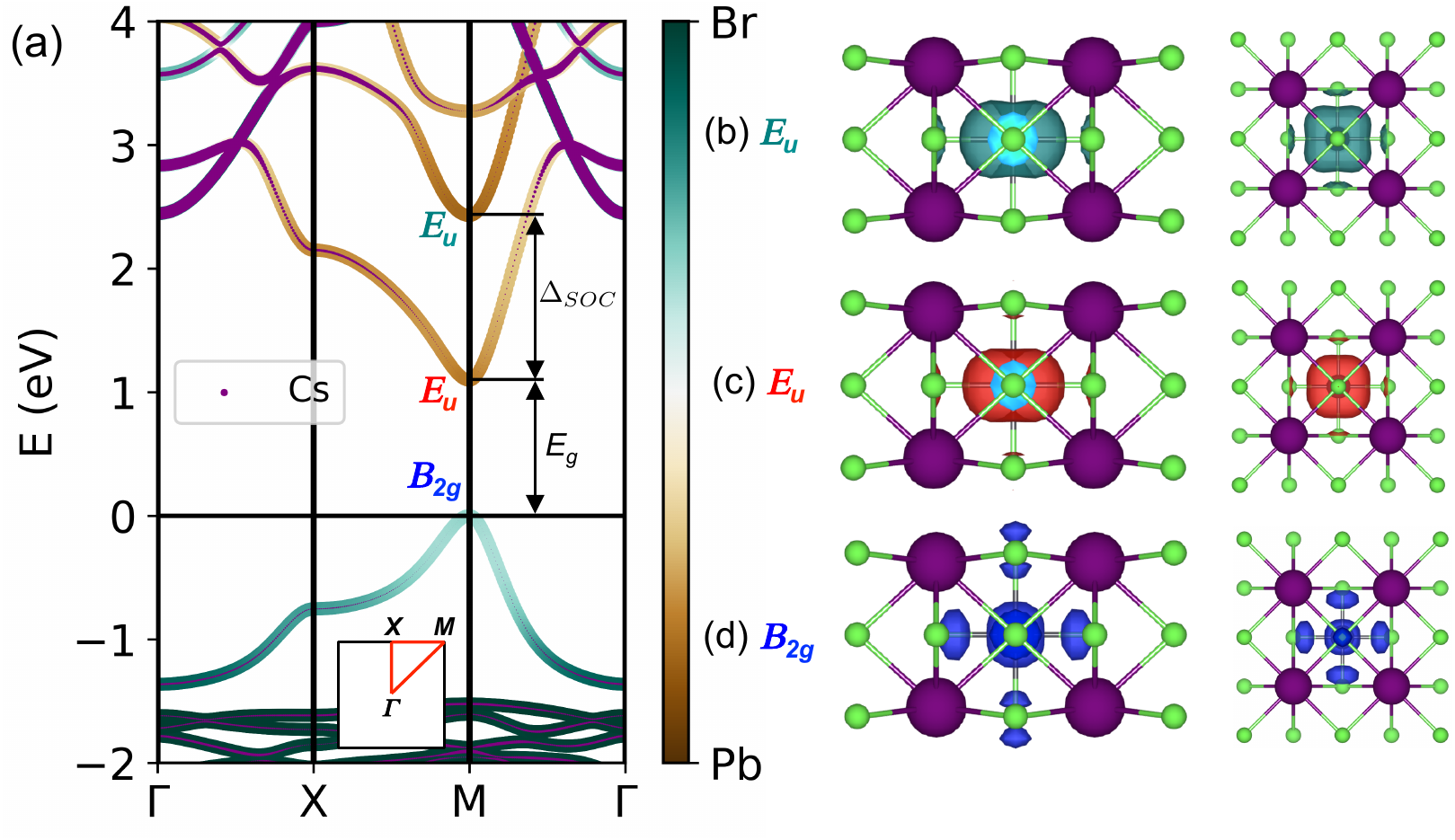}
\caption{(a) Band structure of monolayer Cs$_2$PbBr$_4$. The purple dots are
assigned to the projections onto the cesium atomic orbitals and their size
is proportional to the relative weight. The colormap corresponds to the
relative weight of bromine (green) and lead (brown) atomic orbitals. (b) Wave
functions of the states at $M$: (b) second conduction band $E_u(\Gamma_{6-})$, 
(c) conduction band minimum $E_u(\Gamma_{7-})$, and (d) valence band maximum
$B_{2g}(\Gamma_{7+})$. We use the point group notation instead of the
double-group one.}
\label{bands}
\end{figure*}

We have calculated within the LDA the band structure of monolayer perovskite, as shown
in Fig. \ref{bands}(a). The inclusion of the spin-orbit interaction is fundamental
for a proper description of the band structure of perovskites and therefore we
work with full spinorial wave functions (comparison of the band structure
obtained with
and without the spin-orbit interaction is shown in Fig. S3 of the supporting
information).\cite{Even2013} The DFT bandgap of monolayer 
perovskite is located at the $M$ point with a value of 1.09 eV. Together with the band
structure we have also represented the irreducible 
representations for bulk and monolayer of the valence band and
conduction band states at the bandgap. Figures \ref{bands}(b), (c) and (d) shows
the wave function of the valence and conduction band states at the bandgap at
$M$. The spin-orbit interaction has a sizable effect with a splitting 1.33 eV, even larger than the
DFT bandgap value, which confirms the necessity of using full spinor wave
functions.

In order to make clear the atomic composition of each electronic state, we have 
projected the wave functions onto the atomic
orbitals. In the case of cesium atomic orbitals (purple dots), the dot size
corresponds to the relative weight. Alternatively, we have represented the
relative weight of the projections onto bromine (green) and lead (brown) atomic orbitals
with the color code shown on the color bar, keeping a fixed size. For the
electronic states close to the bandgap the electronic density is localized at
the bromine and lead atoms, as shown by the weights and the wave function
representation, and annotated in Table \ref{table-orbital}. 

Moreover, bands with
relevant weight from cesium atomic orbitals are practically decoupled from the
bands near $M$, placed mainly at $\Gamma$, above 2 eV or at deeper levels around
-13 eV (not shown here). The atomic orbital composition of the states close to the
bandgap has important consequence on the study of the
optical properties. Therefore, from the LDA band structure we conclude that most of the optical activity 
(like absorption or photoluminescence) 
is carried out at the bromine and lead atoms. The cations
(cesium) enter as a stabilizer of the structure but have lesser impact on the
optical properties. Moreover, previous LDA and GW calculations in bulk hybrid perovskites
((CH$_3$NH$_3$)PbI$_3$ and (CH$_3$NH$_3$)SnI$_3$) showed that the 
electronic structure close to the bandgap is barely affected by the molecular
cations.\cite{He2014,Brivio2014} Our calculations of the
electronic structure of bulk (CH$_3$NH$_3$)PbBr$_3$ and CsPbBr$_3$ shows
also that the bands close to the bandgap are very similar in both cases (see
Figs. S4 and S5 of the supporting info). Thus, we can expect that the
substitution of the cesium cations by
molecular cations (like methylammonium) in this 2D material might not change substantially the bands
close to the bandgap and therefore the
conclusions extracted for the optical properties of monolayer Cs$_2$PbBr$_4$ 
can be extrapolated to others semiconducting 2D perovskites like
(CH$_3$NH$_3$)$_2$PbBr$_4$.

\begin{table}
\begin{tabular}{ccccc}
    \hline
    \hline
    Wave function  & IR       &  IR-spin      &  Br  &  Pb    \\
    \hline     
\multirow{2}{*}{CB} & \multirow{2}{*}{E$_{u}$} & $\Gamma_{6-}$ & 13.9 & 82.4 \\
                    &                          & $\Gamma_{7-}$ & 17.9 & 79.6 \\    
\hline
      VB            & B$_{2g}$                 & $\Gamma_{7+}$ & 66.6 & 32.4 \\
\hline
\hline
\end{tabular}
\caption{Relative weight of the projections onto the atomic orbitals
of Br and Pb of the wave functions at the bandgap (valence and conduction
band). The relative weight associated to Cs atoms is negligible. The 
irreducible representations correspond to the single and double space group P4$/$mmm.}
\label{table-orbital}
\end{table}

In addition we have corrected the inherent bandgap underestimation
obtained by the DFT-LDA calculations. We have corrected the LDA eigenvalues
using the GW method, as implemented in the Yambo
code.\cite{Marini2009,Galvani2016,Sangalli2018} The bandgap correction of G$_0$W$_0$ is
1.61 eV resulting in an electronic bandgap of 2.71 eV. The GW correction of the 
conduction bands is basically a rigid shift. In the
case of the valence band states the GW correction shows a deviation with
respect to the rigid shift correction. Therefore, the GW correction
has to be applied to all
the electronic states considered to calculate the optical properties. (see 
full GW band structure in Fig. S1 of the supporting information).
There are alternatives approaches to the use of GW approximation in perovskites.
For instance, the  DFT$-1/2$ method can calculate the electronic structure with
similar accuracy and is computationally more efficient \cite{Tao2017}. Nevertheless,
the goal of this work is to obtain the excitonic states. The BSE and the GW
method are both formulated within the same approach, the many-body perturbation
theory, and for the size of our system they are still computationally efficient, accurate and
parameter-free.

\section{Optical Spectra}

Once we have determined the electronic structure of the monolayer Cs$_2$PbBr$_4$ we calculate the optical
spectra including excitonic effects. We use
the Bethe-Salpeter equation (BSE) with full relativistic spin-orbit as 
implemented in the Yambo code.\cite{Marini2009} The electron and hole energy eigenvalues are obtained
from the GW method and we use the wave functions from the LDA results, 
calculated in previous Section.\cite{Marini2009} The 
vacuum distance between periodic images is 18~\angstrom. We use the Coulomb
cutoff technique in the GW and BSE calculations to avoid artificial interaction between periodic images of the
layer (more details of the BSE formulation can be found in the first section of
the supporting
information).\cite{Rozzi2006}

\begin{figure}
\includegraphics[width=7 cm]{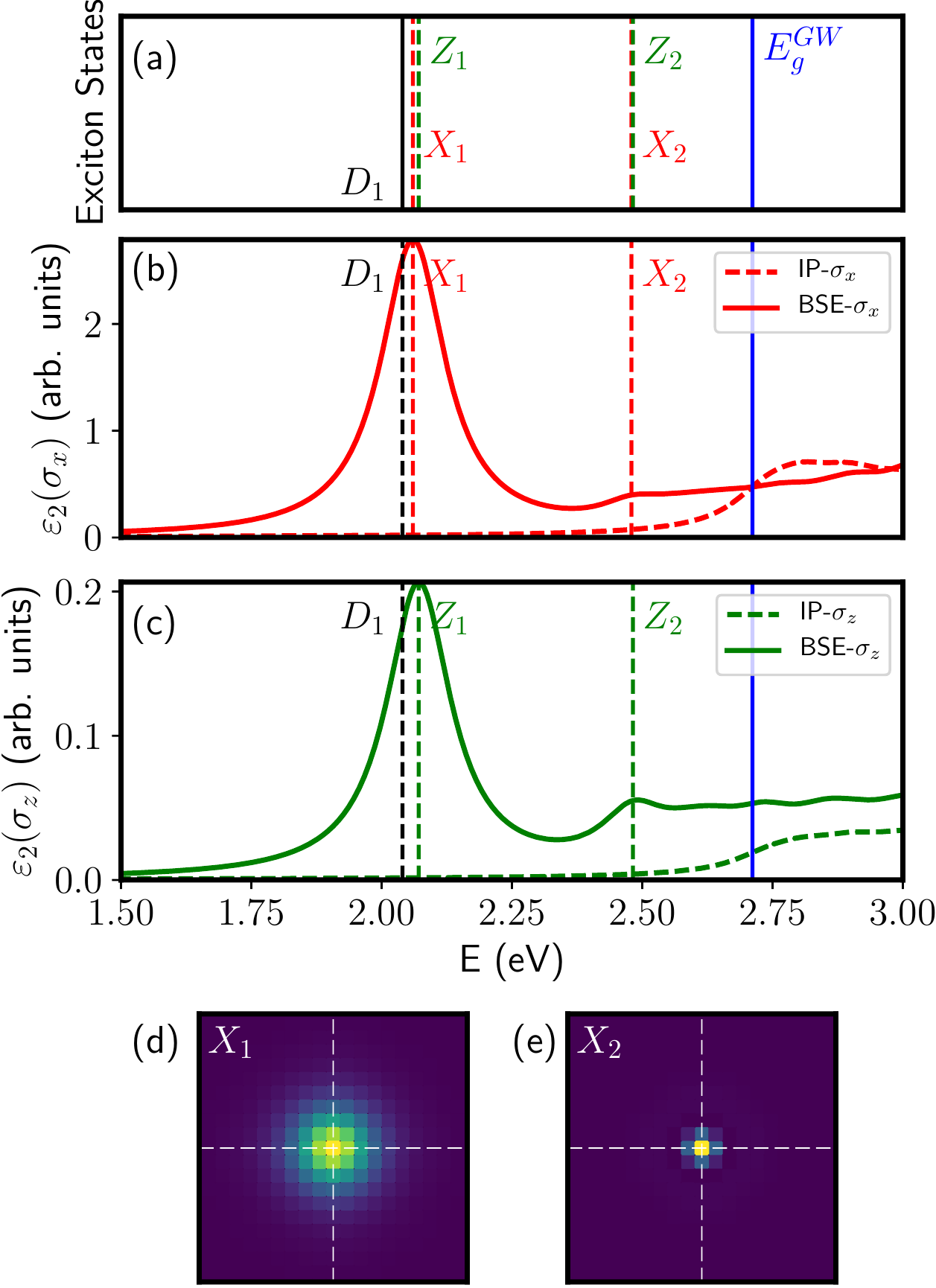}
\caption{(a) Excitonic states and electronic bandgap. Optical spectra for (b) light-polarized along $x$ (in-plane) and
(c) polarized along $z$ (out-of-plane). Independent particle spectra are
represented with dashed lines and BSE spectra with solid lines. The vertical
lines mark the exciton energy positions of the dark exciton (black line),
the bright $X$ and $Z$ series are dotted red and green lines, respectively.
Exciton wave function represented in the 2-dimensional
Brillouin zone of the states (d) $X_1$ and (e) $X_2$.}
\label{bse}
\end{figure}

Figure \ref{bse}(a) shows the main excitonic states. We have represented the
dark exciton $D_0$, the first two bright excitons for light polarized along
$x$ ($X_1$ and $X_2$) and $z$ ($Z_1$ and $Z_2$) directions, together with the 
electronic bandgap, obtained using the GW approximation. The exciton binding
energies for the three lowest states are very similar (see Table
\ref{table-exc}). The dark exciton $D_0$ is at 2.04 eV, 20 meV below $X_1$ (located at 2.06 eV) 
and 31 meV below $Z_1$ (located at 2.071 eV). The excited state excitons $X_2$ and $Z_2$
exhibit a smaller binding energy in comparison with $X_1$ and $Z_1$ (0.22 eV).
The binding energies of 2D perovskites are much larger than in
the case of the bulk. The BSE calculations for bulk CsPbBr$_3$ give a binding energy
of 68 meV (see BSE spectra in Fig. S6 of the supporting information). The
exciton binding energy of bulk is in agreement with previous \textit{ab initio} BSE
calculations performed for CH$_3$NH$_3$PbBr$_3$, which obtained an exciton binding energy of 71
meV\cite{Bokdam2016}.

In order to verify the composition and symmetry of the excitonic states we have
represented the exciton wave functions in the reciprocal $\bf k$-space (see
first section of the supporting information 
for definitions). Figure \ref{bse}(d) and (e) shows $w_{\bf k}$ for
excitons $X_1$ and $X_2$. The weights of excitons $D_0$ and $Z_1$ 
are very similar to the one of $X_1$ and they are not represented. In all the cases
the weight is localized around the $M$ point, at the direct bandgap. The
excitonic weights confirm that the optical activity is carried out by wave
functions localized at the bromine and lead atom, in agreement with the
calculations of the electronic structure. Therefore, in a first approximation,
we can neglect the role of the cations, either organic or inorganic, to
delve into the optical properties of monolayer perovskites.

The absorption spectra gives a more complete picture of the optical properties. Figures 
\ref{bse}(b) and (c) shows the optical spectra of monolayer Cs$_2$PbBr$_4$,
obtained at the independent particle approximation (IP, dashed lines) and
at the BSE level (solid lines). We represent the absorption for two
configurations of light polarization, (b) along $x$ (lines in red) and (c) along $z$ 
(lines in green). The spectra are obtained by choosing a Lorentzian width of 75 meV. As expected, the 
intensity of the spectra $\epsilon(\sigma_x)$ is much  larger than the one of
the $\epsilon(\sigma_z)$ due to the depolarization effects.\cite{Wirtz2006} In both
cases we have a strong renormalization of the IP spectra due to the excitonic
effects and the brightest peaks are associated to the $X_1$ and $Z_1$ excitons.

\begin{figure*}
\includegraphics[width=14 cm]{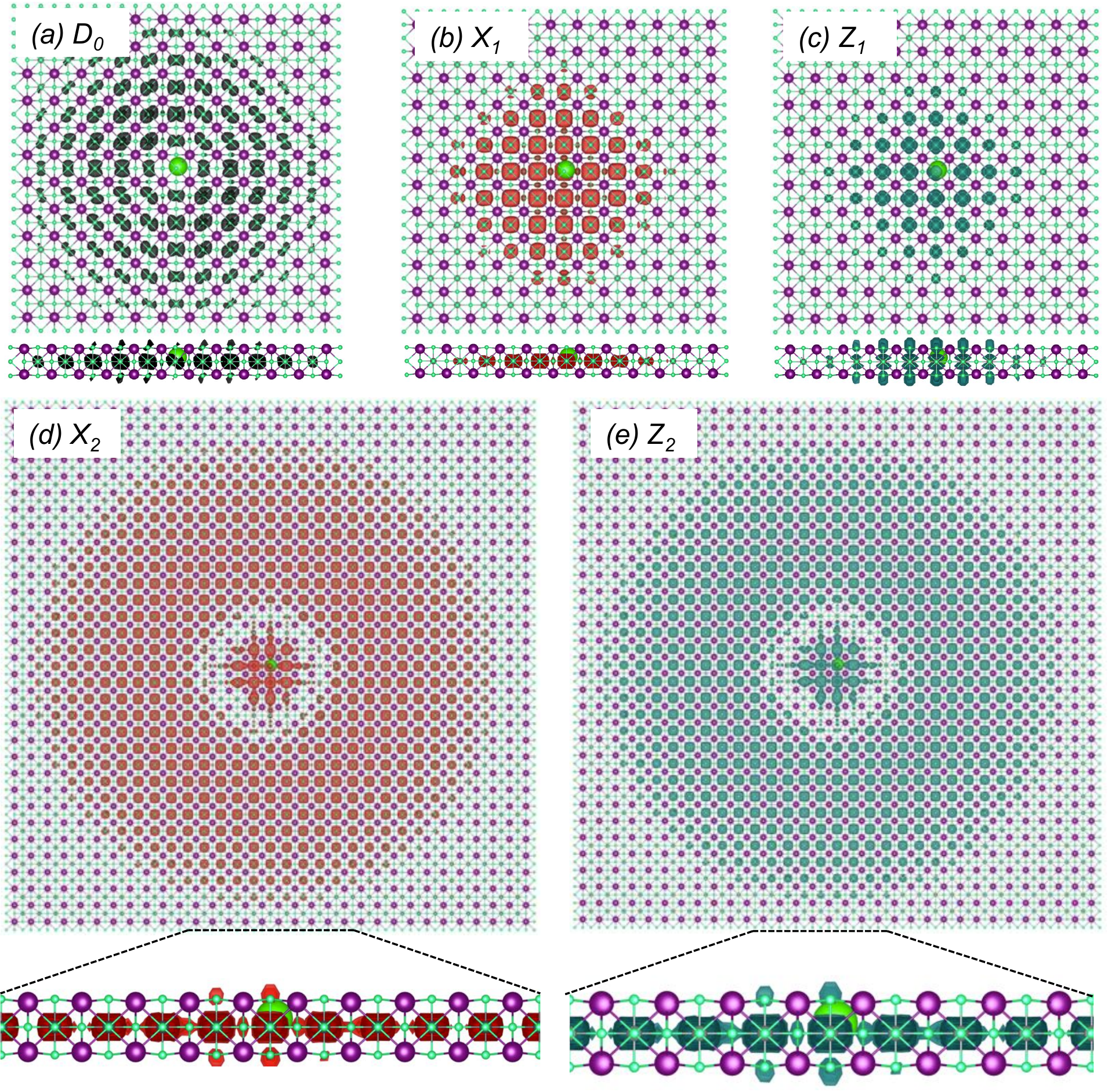}
\caption{Exciton wave functions in real-space of the states (a) $D_0$,
(b) $X_1$, (c) $Z_1$, (d) $X_2$ and (e) $Z_2$ (top and lateral view). The 
green sphere represents the position of the hole (size exaggerated for better
visibility).}
\label{wf}
\end{figure*}

In addition to the $\bf k$-space plots, the real space representation gives further
information regarding the localization and symmetry of the excitons. As we have
mentioned, only from the $\bf k$-space plots for excitons $D_0$, $X_1$ and $Z_1$
is difficult to highlight the differences between these states. Figure
\ref{wf}(a), (b) and (c) shows the
electronic density of the states $D_0$, $X_1$ and $Z_1$, respectively. In order
to analyze the exciton symmetry, we have
placed the hole near the bromine atom. From the symmetry analysis of the
excitons wave functions we assign the $A_{2g}$ representation to the $D_0$
exciton. Therefore, $D_0$ does not couple to either in-plane light ($E_u$) 
or out-of-plane light ($A_{2u}$). Another feature of the $D_0$ excitons is that
electronic density is spatially separated from the hole density. This is
evidenced by performing BSE calculations without exchange interaction which does
not modify the exciton binding energy.

The bright excitons $X_1$ and $Z_1$ have $E_u$ and $A_{2u}$
and the selection rules impose coupling to in-plane and out-plane light,
respectively. The selection rules can also be deduced from the inspection of 
their wave functions (in this particular case). If we place
the hole near the lead atom, we observe that the density
around bromine atoms resemble the shape of atomic orbitals with $p_x+p_y$
symmetry for the $X_1$ exciton and with $p_z$ symmetry for the $Z_1$ exciton.
Moreover, we have represented the excited exciton states $X_2$ and $Z_2$ in Fig.
\ref{wf}(d) and (e). These states are much less localized than the first three
excitons in accordance with the smaller oscillator strength. They follow the
same selection rules that their counterparts $X_1$ and $Z_2$.

\begin{table}
\begin{tabular}{l|ccccc}
    \hline
        State       &  $D_0$   &  $X_1$    & $Z_1$   & $X_2$   & $Z_2$   \\
    \hline
        Energy (eV) &  2.040  &  2.060  & 2.071 & 2.479 & 2.482  \\
    \hline
Binding Energy (eV) &  0.672  &  0.652  & 0.641 & 0.233 & 0.229  \\
    \hline
\end{tabular}
\caption{Excitonic energies and binding energies. The binding energy is defined
with respect to the GW bandgap ($E_{g,GW}-E_{exc}$).}
\label{table-exc}
\end{table}

Our calculations provide some results to be validated by optical experiments.
For example, the impact of the dark exciton $D_0$ on the optical properties could be assessed
by performing PL as a function of temperature. As summarized in Table
\ref{table-exc}, the energy separation between $D_0$ and $X_1$ is 20 meV. Thus, increasing 
population of the $D_0$ exciton in detriment of $X_1$
exciton (by reducing temperature) can affect the PL efficiency. The PL
kinetics would also change with temperature because the radiative lifetimes
are determined by the exciton oscillator strength and they are different by several order
of magnitude between $D_0$ and $X_1$ excitons.\cite{Palummo2015,Chen2018} Moreover, the 
electron-hole separation observed in the wave function
of $D_0$ in Fig. \ref{wf} together with the slow radiative recombination can
also play an important role in efficient photoinduced charge
separation.\cite{Liu2017} At this respect, 
magentophotoluminescence experiments can help to elucidate the splitting between $X_1$ and $Z_1$
excitons, or the nature of the dark exciton $D_0$.\cite{Robert2017}


Moreover, our theoretical results is useful to interpret the optical
experiments available for organic 2D
perovskites of 
(CH$_3$NH$_3$)$_{n+1}$Pb$_n$Br$_{3n+1}$ and
(CH$_3$NH$_3$)$_{n+1}$Pb$_n$I$_{3n+1}$.\cite{Blancon2018,Blancon2018a,Smith2017a} 
The reported values of the exciton binding
energy are 467 meV
and 490 meV (monolayer (C$_4$H$_9$NH$_3$)$_2$PbI$_4$), 
against the 652 meV of our
calculations.\cite{Yaffe2015,Smith2017,Blancon2018,Blancon2018a} In both cases the large
value indicates strong excitonic effects, typical of 2D semiconductors. The
difference in the value comes from the different dielectric screening. We assume
a free-standing monolayer for the calculations while the system of Ref.
\cite{Blancon2018a} is surrounded by organic molecules acting as spacers, increasing the
dielectric screening and therefore reducing the exciton binding energy.\cite{Qiu2017} Other
causes for the deviation are the
different bandgap of both systems (1.90 eV in
organic perovskite against the 2.71 eV of monolayer perovskite), which
can modify the exciton binding energy\cite{Choi2015,Jiang2017}  (a comparison
with experimental data is shown in Fig. S7 of supporting info). The 
agreement obtained by the semi-empirical BSE approach developed by
Blancon et. al.\cite{Blancon2018a}
is remarkable, in part thanks to the parabolic dispersion of bands near the
bandgap. Nevertheless, the spin-orbit interaction is not included
and in systems with a more complicated band structure this model can lose some accuracy.


The relatively simple synthesis of perovskites materials makes possible the
fabrication of inorganic 2D materials varying composition. For instance, the
replacement of lead by tin and bromine by iodine or chlorine opens the way to
change the bandgap within 1 eV,\cite{Huang2013} and consequently the exciton
binding energy. Moreover, the synthesis of nanoplatelets point towards a
feasible manipulation of the number of layers.\cite{Weidman2017}
In future works, we plan to study the optical properties of 2D
perovskites for different chemical compositions and as a function of the number
of layers.
The synthesis of 2D inorganic perovskites together with the use of molecule as
spacers can make possible the growth of colloidal supercells of single-layer
perovskites. This single-layer supercell would have a larger
optical efficiency due to the larger quantity of active material.\cite{Rodriguez-Romero2017} 

\section{Conclusions}

Our approach allows the application of state-of-the-art \textit{ab initio} simulations 
like the Bethe-Salpeter equation and the GW method in a simplified model 2D material. 
We have found a stable 2D perovskite, monolayer Cs$_2$PbBr$_4$, with remarkable
excitonic and optical properties. The monolayer shows a direct bandgap at the
$M$ point and a giant spin-orbit coupling. We have found the excitonic states and classified them in
dark and bright excitons. The excitons exhibit a strong exciton 
binding energy (0.652 eV for the bright exciton).

From the theoretical point of view, 
simulations in this kind of 2D perovskites are useful to understand the
properties of more complex organic 2D perovskites. This system is suitable for more complex simulations, 
like for instance the study of the electron-phonon
interaction to explain non-radiative recombination,\cite{Guo2016} 
charge recombination rates\cite{Huang2017,Chen2018}, biexctions\cite{Thouin2018},
high-harmonic generation,\cite{Abdelwahab2018} and carrier
kinetics.\cite{Liu2017,Molina-Sanchez2017}  So far the synthesis of monolayer Cs$_2$PbBr$_4$
has not been reported but we believe that our work will stimulate efforts
addressed toward the
synthesis of inorganic 2D perovskites.\cite{Rui2017} Future works will deal with \textit{ab initio} approaches to
simulate the optical properties of organic 2D perovskites and their dependence
on the number of layers.

\section{Associated Content}
Supporting Information Available: Details of the GW and BSE calculations, phonon
band structure, details of the effects of spin-orbit interaction.

\section{Acknowledgements}

I acknowledge the Juan de la Cierva Program (Grant IJCI-2015-25799) of
Spanish Government for its financial support, and my colleagues Juan Mart\'{i}nez-Pastor, Alberto
Garc\'{i}a-Crist\'{o}bal, Ludger Wirtz, Fulvio Paleari, and Jacky Even for the
reading of the manuscript, and their valuable comments and suggestions.

ASSOCIATED CONTENT Supporting Information Available:

\bibliography{biblio}

\end{document}